\documentclass[aps,prl,reprint,superscriptaddress,longbibliography]{revtex4-2}
\usepackage{graphicx}
\usepackage{bm} 
\usepackage{xcolor}
\usepackage{hyperref}
\usepackage{bigints}

\newcommand{\vect}[1]{\mathrm{\mathbf{#1}}} 

\newcommand{\refeq}[1]{Eq.~(\ref{#1})}

\begin{document}

\title{Photocurrent-induced harmonics in nanostructures}

\author{Ihar Babushkin}
\affiliation{Institute of Quantum Optics and PhoenixD, Leibniz
  University, Welfengarten 1, 30167 Hannover, Germany}
\email{babushkin@iqo.uni-hannover.de}
\affiliation{Cluster of
  Excellence PhoenixD (Photonics, Optics, and Engineering-Innovation
  Across Disciplines), 30167 Hannover, Germany}
\affiliation{Max Born Institute, Max Born Str. 2a, 12489 Berlin,  Germany}

\author{A. Husakou}

\affiliation{Max Born Institute, Max Born Str. 2a, 12489 Berlin,  Germany}

\author{L. Shi}

\affiliation{Hangzhou Institute of Technology, Xidian University,
  Hangzhou 311200, China}
\affiliation{School of Optoelectronic
  Engineering, Xidian University, Xi’an 710071, China}

\author{A. Demircan}

\affiliation{Institute of Quantum Optics and PhoenixD, Leibniz
  University, Welfengarten 1, 30167 Hannover, Germany}
\affiliation{Cluster of
  Excellence PhoenixD (Photonics, Optics, and Engineering-Innovation
  Across Disciplines), 30167 Hannover, Germany}

\author{M. Kovacev}

\affiliation{Institute of Quantum Optics and PhoenixD, Leibniz
  University, Welfengarten 1, 30167 Hannover, Germany}
\affiliation{Cluster of
  Excellence PhoenixD (Photonics, Optics, and Engineering-Innovation
  Across Disciplines), 30167 Hannover, Germany}

\author{U. Morgner}

\affiliation{Institute of Quantum Optics and PhoenixD, Leibniz
  University, Welfengarten 1, 30167 Hannover, Germany}
\affiliation{Cluster of
  Excellence PhoenixD (Photonics, Optics, and Engineering-Innovation
  Across Disciplines), 30167 Hannover, Germany}

 \begin{abstract}
 Photocurrent-induced harmonics appear in gases and solids
  due to tunnel ionization of electrons in strong fields and
  subsequent acceleration. In contrast to three-step harmonic emission, no return to the parent ions is
  necessary. Here we show that the same mechanism produces
  harmonics in metallic 
  nanostructures in strong fields. Furthermore, we demonstrate how strong local field
  gradient, appearing as a consequence of the field enhancement, affects
  photocurrent-induced harmonics. This influence can shed light at
  the state of electron as it appears in the
  continuum, in particular, to its initial velocity.
   \end{abstract}

\date{\today}

\maketitle

\section{Introduction}

In strong optical fields, electrons leave the surface of metallic
nanostructures (NS) -- the process in many respects similar to
photo-induced ionization of atoms
\cite{vampa17,dombi20,kruger2011-nanotip-tunnnel,kruger12-metal-nanotip-rev}.
Short, few-cycle optical pulses allow to induce sub-cycle
tunneling dynamics of electrons leaving NSs \cite{kruger2011-nanotip-tunnnel,kruger12-metal-nanotip-rev,dombi13,dombi20,ludwig2019sub,shi21}. In the last decade this process has attracted strong and growing interest in context of attosecond
science. Most of the
attention is paid to the dynamics of electrons themselves, which is of high importance in the context of generation of on-chip petahertz
electronics
\cite{vampa17,dombi20,schoetz2019perspective,shi21,schiffrin13,karnetzky18}, scanning electron microscopy in near-field emission mode \cite{sem}, and emission directly by the plasmonic fields \cite{plasm_em}. 
Equally important are attempts to generate high harmonics (HHG) in the extreme ultraviolet range created by electrons
returning back to the NS
\cite{ciappina14} or to the atoms in vicinity of
  NSs \cite{kim08,dombi20}.

In this article we consider, in contrast, a photoinduced-current mechanism. The electrons which emerge in the continuum and
are subsequently accelerated by the field also emit radiation, which does not depend on their return to the parent ion
\cite{brunel90cp,kim08b,babushkin11,lanin17,babushkin17,jurgens20,babushkin22}.
This radiation, typically located at lower frequencies than HHG,
attracted much less attention in the strong field optics, with an exception of the lowest-order (0th) harmonic. The latter is typically located
in terahertz (THz) range, with photoinduced current providing a very
efficient mechanism for its generation \cite{kim08b,zhang17,koulouklidis20}. 
The higher-order
photoinduced current harmonics attracted significant attention only
recently \cite{lanin17,babushkin17,jurgens20,babushkin22}. In
particular, it was shown \cite{babushkin22} that such harmonics can
provide details of the redistribution of
electronic wavepackets at a deeply-subcycle scale, even if harmonic wavelength is much larger than
the atomic scale, as well as details about the electron dynamics in the crystal lattice \cite{jurgens24}. While in the context of photoinduced current mostly
the Brunel mechanism \cite{brunel90cp,babushkin17} is
considered, related to the change of the refractive index as the
plasma is created, recent studies demonstrated that the creation
of the free charge itself \cite{jurgens20} (accomplished by absorbing energy
from the driving pulse) is also responsible for emission of harmonics by the so-called injection current.

Here, we study the emission of both Brunel and injection-current
harmonics in metallic NSs at THz and
high frequencies. We predict that the same mechanisms which acts in
gases and solids will also create the photocurrent-based harmonics
in the case of NSs. We show that the strong field
gradient can significantly reduce the harmonic emission efficiency in the case if
the gradient is large enough. Finally, we show that the scaling of
harmonic energy with the field gradient can shed light on the
dynamics of the electron wavepacket at the exit of the tunneling
barrier. 

The paper is structured as follows: in Section 2, we establish the semiclassical model and derive the expression for the emitted harmonics. In Section 3, we analyze the effect of the field gradient on the harmonic emission in a simplified case. In Section 4, the result of full quantum time-dependent Schr\"odinger equation (TDSE) simulation are presented, followed by a conclusion. 

\section{The system}

Here we consider nanostructures, irradiated by strong
  few-cycle pulses [see inset in Fig. 1b]. The few-cycle
  duration of the pulses allows to achieve the tunnel-like
  photoemission process [see inset in Fig. 1(a)] without immediately  destroying the nanostructures \cite{vampa17,dombi20,shi21}. Even if
  the field intensity in the vicinity of the nanostructure can
  approach several tens of TW/cm$^2$, NSs in the few-cycle regime 
  can still sustain milliards of pulses \cite{shi20}. We focus
  specifically on metallic NSs, although our results are in general
  applicable to the dielectric NSs. The important feature
  of metallic NSs with sharp edges and features is the strong field
  enhancement near these sharp features. The enhancement mechanism is
  related to fast charge redistribution inside the NS and to
  excitation of plasmons. The field enhancement in the vicinity of the
  NS is localized to a small spatial region, which is equivalent to
  strong field gradients near the surface.

Here, we assume for simplicity a small spatial region near
  the NS surface given by $x=0$ [see inset in Fig. 1(b)]. We consider a
  one-dimensional geometry with all quantities depending on the
  spatial coordinate $x$. The pump field is polarized in $x$ direction
  with amplitude described by
\begin{equation}
  \label{eq:field-grad}
  E(t,x) = E_0(t) \left(1 + (f-1)e^{-\alpha x}\right)
\end{equation}
for $x>0$ (outside the NS), where the constant $\alpha$ describes the field gradient while $f$
describes the field enhancement factor. 
Note that field polarization corresponds to the realistic situation, since the boundary conditions prevent the existence of the tangential field at the surface of an (ideal) metal. The exponential behavior of the enhancement corresponds to the spatial structure of an evanescent wave of a plasmon, which decays exponentially with the distance from the metal surface.  Equation
\refeq{eq:field-grad} is formulated in such a way that for $x\to
\infty$ we have $E(t)\to E_0(t)$, whereas for $x\to 0$ we have
$E(t)\to fE_0(t)$. 

\section{Semiclassical model}
To gain an intuitive understanding of the relevant processes, we start from a semiclassical model, which will be followed in Section 4 by a full quantum simulation.
The driving field $\vect E(t)$ leads to tunneling of electrons from the
  surface to the continuum [see inset in Fig. 1(a)] and
subsequent acceleration. A continuum electron (with charge $-q$) creates a dipole $\vect P = -q\vect r$ where $\vect r$ is its position. In addition, in principle, it redistributes the electrons at the metallic NS
surface, which can be described by a single or multiple effective "image" electrons. In the simplest case of flat ideal-metal surface, this would result in doubling the dipole. However, in realistic case, the value and position of image electron(s) depend on the geometry and dielectric function of the NS, on top of that, the redistribution has a finite response time. Therefore in this manuscript we prefer to disregard the image charges; for a specific NS shape and dielectric function, they can be easily taken into account by multiplying the dipole by a corresponding factor. 

\begin{figure}
\includegraphics[width=\columnwidth]{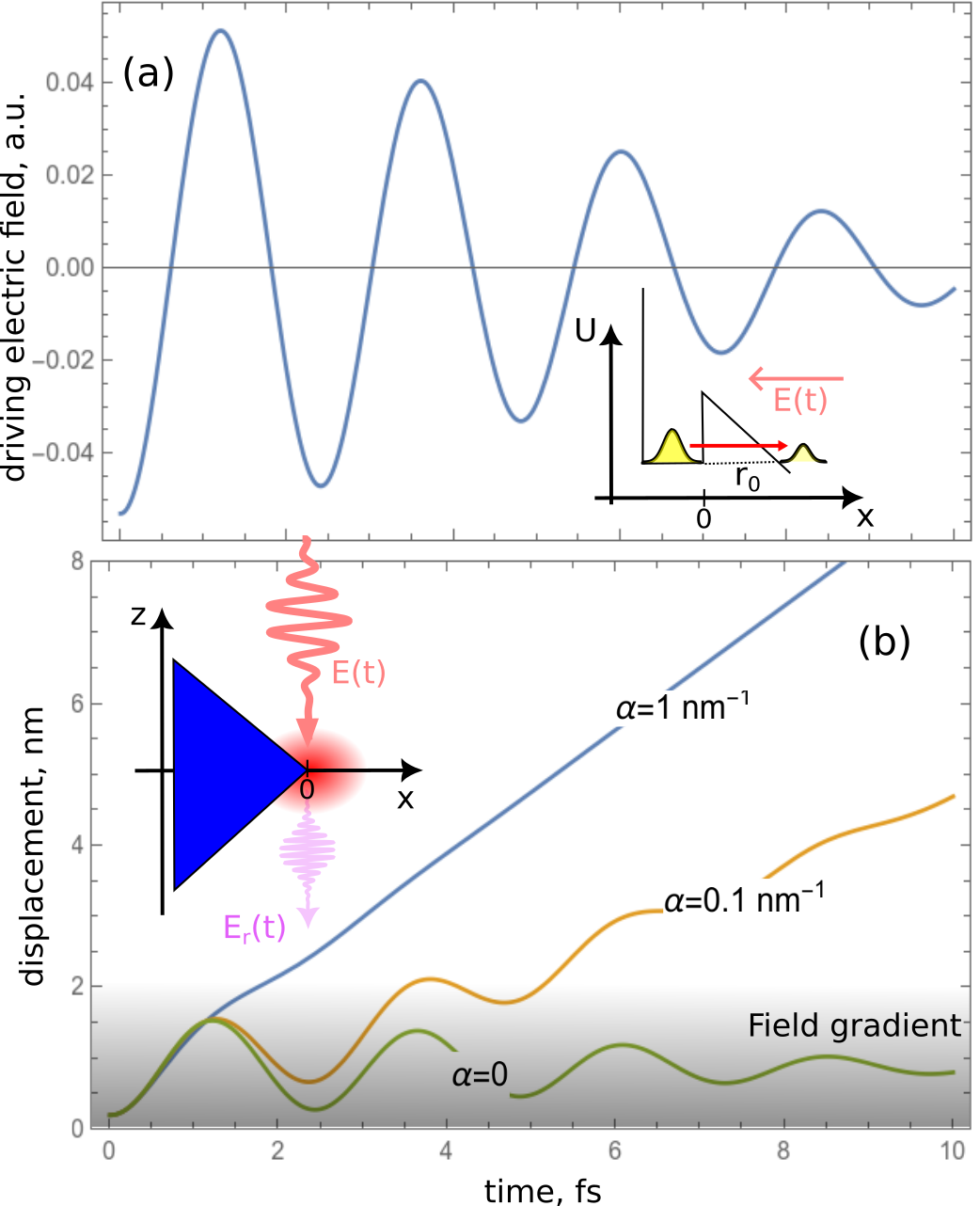}
  \caption{(a) An exemplary strong few-cycle pump pulse $\vect E(t)$
    (only part of the pulse for $t>0$ is shown), acting on an
    asymmetric metallic NS [see inset in (b)]. This
    is modeled (see inset) using single-electron in an asymmetric
    potential (thin black line) deformed by the external electric
    field, resulting in tunneling of the electronic wavepacket (yellow
    shapes, red arrow). (b) Exemplary trajectories of electrons,
    created at $t=0$ near the maximum of electric field, for different
    values of the enhanced field gradient $\alpha$. The inset to (b)
    shows the excitation geometry:  The driving field $\vect E(t)$ leads to  the field enhancement (red region) near the tip of the nanostructure (blue rectangle) and to emission of electrons
     and radiating nonlinear
    response $\vect E_r(t)$. It is assumed that all the fields are
    polarized in x-direction, and the movement of electrons happen
    also in the same direction [see inset in (a)].
  \label{fig:base}}
\end{figure}


In the classical approximation we can describe the velocity  $
\vect v(t,t')$ of a nonrelativistic electron at the
time $t$, born at time $t'$, as
\begin{equation}
  \label{eq:1el}
  m\frac{d\vect v(t,t')}{dt} = -q\Theta(t-t')\vect E(t,\vect r),
\end{equation}
where $m$ is the electron mass, $\vect E(t,\vect r)$ is the
driving electric field at time $t$ and position $\vect r$. $\Theta(t)$
is the Heaviside step-function, describing the appearance of the electron in
the continuum. The moving electron
radiates the field $\vect E_r\propto d\vect v/dt$.  
\refeq{eq:newton} can be integrated, giving
\begin{equation}
  \label{eq:newton}
\vect v(t,t') = \vect v^{(0)} -
\frac{q}{m}\int_{t'}^t \vect E(\tau)d\tau,
\end{equation}
where $\vect v^{(0)}$ is the initial velocity of the electron at time
$t'$.

The change $d\vect P$
of the total dipole $\vect P = -q\sum_i\vect r_i $, obtained by summation over all
the dipoles in the continuum, is given (after approximation
of the summation by integration) by
\begin{equation}
  \label{eq:dP1}
  d\vect P(t)  = -q \left(\dot \rho(t) \vect r^{(0)}   +
  \int \dot \rho(t')\vect  v(t,t')dt' \right)dt,
\end{equation}
where $\vect r^{(0)}$ describes the distance from the surface at which the charges emerge
in the continuum [see inset in Fig. 1(a)], $\rho(t)$ is the density of ionized electrons and $\dot \rho(t)$ is the ionization rate. 

The first term in the right hand side of
\refeq{eq:dP1} describes the change of the dipole due to creation of
new electrons in the continuum, whereas the second term is responsible
to the modification of the dipole due to acceleration of already
existing dipoles (created at earlier times $t'$). For the moment being,
we have neglected the term with $\vect v^{(0)}$ in \refeq{eq:dP1}.

Assuming  $\vect v(t,t')$ given by \refeq{eq:newton}, and
differentiating  \refeq{eq:dP1}, we obtain the current  $\vect J=-q\sum_i\vect v_i = d\vect P/dt$ as
\begin{equation}
  \label{eq:dj}
  \frac{\partial\vect J}{\partial t} =
  -q\frac{\partial}{\partial t}\left({\vect r^{(0)} \dot
      \rho}\right)  + \frac{q^2}{m}\vect E\rho.
\end{equation}
 Finally, the radiation $E_r$,  observed by a remote observer, is
\begin{equation}
  \label{eq:Er}
  \vect E_r = g\frac{\partial{\vect J}}{\partial t},
\end{equation}
where $g\propto 1/R$ is a geometrical factor, depending on the
distance $R$ to the detector.

Expression \refeq{eq:dj} is analogous to that for gases and solids. The last contribution in
\refeq{eq:dj} is known as Brunel mechanism and describes the
contribution to the radiation [\refeq{eq:Er}] due to acceleration of
already existing dipoles, whereas the former term in the right hand
side of \refeq{eq:dj} is known as injection current and describes the
creation of the dipoles in the continuum during tunneling. Note that
if no electrons are photoionized ($\dot \rho=0$), the first term in
\refeq{eq:dj} is zero and the second term contributes only to the
modification of pump field phase without emission of new spectral components, since $\rho(t)=\mathrm{const}$. On the other hand,
$\dot \rho\ne 0$ leads to emission of harmonics. In addition, here we consider
asymmetric structures, so both even and odd harmonics of the driving
field can be created, in contrast to gases.

\section{Impact of Field Gradients}

\begin{figure}[tbph!]
  \centering
  \includegraphics[width=\columnwidth]{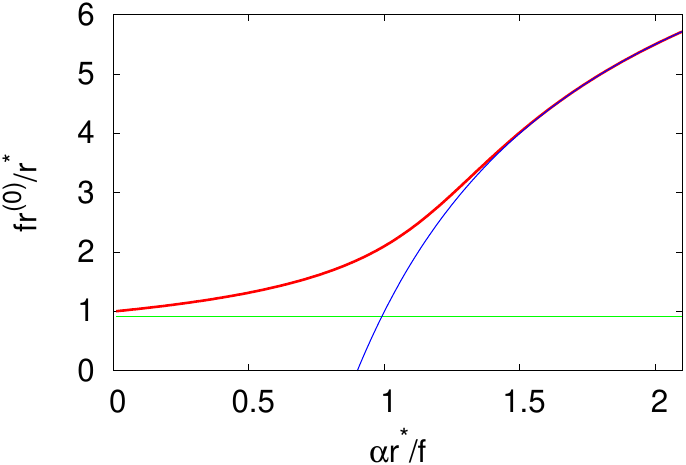}
  \caption{The dependence of the tunnel exit $r^{(0)}$ on the field gradient $\alpha$. The tunnel exit (red curve) was calculated for $f=10$. By green line, the value $r^{(0)}=r^*/f$ is given, while by blue curve, the function $r^{(0)}=r^*-(f-1)/\alpha$ is shown. }
  \label{fig:exit}
\end{figure}

The tunnel exit determines the strength of the injection-current
emission, which was experimentally shown to be the dominant mechanism
of the low-order harmonic generation in several bulk crystals
\cite{jurgens20,jurgens24}. It is therefore instructive and important
to analyze the dependence of this quantity on the gradient of the
enhanced electric field. From here on we consider linearly polarized
electric field and the induced one-dimensional geometry (see insets to
Fig.1), therefore for all vectorial quantities we assume the x-direction
and use the corresponding scalar amplitudes. The tunnel exit occurs at distance $r^{(0)}$ from the NS surface in the direction of the electric field, with $r^{(0)}$ determined from the condition
\begin{equation}
-q\int_0^{r^{(0)}}E(t,r)dr=U,
\end{equation}
where $U$ is the work function (i.e. energy necessary to remove a single electron from NS). In homogeneous field this condition trivially yields $r^{(0)}=r^*\equiv-U/[E(t)q]$, equivalent to the known tunnel exit expression for isolated atoms and molecules. In a non-homogeneous fields, such as the one given by \refeq{eq:field-grad}, this condition is modified. In this case the tunnel exit is given by the transcendental equation 

\begin{equation}
\frac{r^{(0)}}{r^*}+(f-1)\frac{1-e^{-\alpha r^{(0)}}}{\alpha r^*}=1
\end{equation}
which was numerically solved as shown in Fig. \ref{fig:exit}. One can see that for small inhomogeneity, the tunnel exit (red curve) is close to the value $r^{(0)}=r^*/f$ (green line), since in this case the electron feels an almost homogeneous enhanced field $fE(t)$. On the other hand, for large inhomogeneity, the tunnel exit approaches the value $r^{(0)}=r^*-(f-1)/\alpha$ (blue curve) which corresponds to the situation when electron, after passing through a relatively thin layer of enhanced field $fE(t)$, tunnels far into the space region with non-enhanced field $E(t)$. Thus we predict the transition from enhanced-field to non-enhanced-field regime in the generation of the injection current, dependent on the gradient.

We note that for conditions such as intensity of 100 TW/cm$^2$ and $U$ of several eV, the tunnel exit is several angstroms, which is less than the inhomogeneity scale which is typically few nanometers. Tunnel exit can reach larger values for high work function and weak pump fields; however, it is possible that in this situation the tunneling rate is small, making tunneling exit less relevant.

Let us turn our attention to electron dynamics immediately after photoionization. Some exemplary electron trajectories for different typical values of
$\alpha$ are shown in Fig. 1(b) for the 7-fs, 100-TW/cm$^2$ Gaussian pulse at
frequency $\omega_0$, corresponding to the central
wavelength of 800 nm [see Fig.
1(a) (only the region of positive times is shown)], and gold NS with $U=5.1$ eV. We note parenthetically that the intensity of 100 TW/cm$^2$ refers to the enhanced field outside the NS; the intensity inside the NS will be much lower due to boundary condition, which in conjunction with the short pulse duration mean that we can avoid the damage for NS materials such as gold. One can see that the
gradient significantly modifies the electron trajectories. Not only the final electron velocity but also its acceleration strongly depend on $\alpha$. In addition, the tunnel exit is visible at $t=0$, however, it is almost independent on $\alpha$. Since the
radiation $E_r$ is proportional to electron acceleration, we can expect
also the modification of the photoinduced radiation by the field
gradients near the NS. 

One can try to solve the semiclassical model by substituting
$\vect E(t,\vect r)$ in \refeq{eq:1el} by \refeq{eq:field-grad} and
integrating for different electron trajectories. Later on, instead of
making such simulations, direct simulations of the quantum electron
dynamics is performed (see the next section).

On the other hand, the
semiclassical model can be further simplified to give us purely
analytical insight into the dynamics and emission scaling. For this, we replace $\exp(-\alpha x)$ in
\refeq{eq:field-grad} by  a piece-wise approximation
described as
\begin{equation}
E(t,x)=E(t)\begin{cases}
      f,&\text{if $x<\alpha^{-1}$;}\\
      1,&\text{if $x>\alpha^{-1}$}.
    \end{cases}
  \label{eq:e-simpl}  
\end{equation}

The validity of this approximation depends on the relation between, on one hand, the maximum excursion of the electron (twice the amplitude of the spatial oscillation in the field $E(t)=fE_0\cos(\omega_0 t)$) given by 
\begin{equation}
  \label{eq:xmax}
  x_\mathrm{max} = \frac{2qfE_0}{m\omega_0^2},
\end{equation}
and on the other hand the spatial scale of gradient $\alpha^{-1}$.
The relation between $x_\mathrm{max}$ and $\alpha^{-1}$
  determines how the electron “feels” the field gradient; namely, if $x_\mathrm{max}\ll \alpha^{-1}$, the electron will not enter the low-field area and will not recognize any effect of gradient.
We note that the dimensionless parameter $\delta$ introduced in \cite{kiss22} has a
similar role and is related to our parameters by $\delta = 2/(\alpha x_\mathrm{max})$.
 
Let us estimate the typical values of
these quantities. As can be seen from Fig. 1(b), for the parameters
considered in this section $x_\mathrm{max}$ is around 1.5 nm. However,
it can be significantly larger for stronger fields at longer
wavelengths, for example, it would reach 8 nm for 2400-nm pulse with
the intensity (after enhancement) of 300 TW/cm$^2$. On the other hand,
the scale of inhomogeneity $\alpha^{-1}$ is typically well below the
curvature radius of the sharp features of the NS \cite{hhg_np} (or in
a case of a nanoparticle, below its radius). In turn, these radii can
in many cases reach the sub-10-nm values, which supports the validity
of the stepwise approximation for practical systems. For
$x_\mathrm{max}< 1/\alpha$ the electron will always ``feel'' the
homogeneous field, however, as we will see later, in this case the
inhomogeneity plays only a minor role anyway.

Let us estimate the time $t''$ which is required for the electron to reach $x=\alpha^{-1}$ in the case $x_\mathrm{max}\gg 1/\alpha$. For the evolution of electron immediately after the ionization, which typically happens at the maximum of the electric field $E(t)=-fF_0$, we use second Newton's law to write 

\begin{equation}
x(t)=v^{(0)}t+\frac{fqF_0t^2}{2m}.
\end{equation}

While an analytic expression for the initial electron displacement (tunnel exit) $r^{(0)}$ is readily available, the value of $v^{(0)}$ is harder to estimate. For large $v^{(0)}$, we get 
\begin{equation}
  \label{eq:t}
  t''=\frac{1}{v^{(0)}\alpha},
\end{equation}

while if $v^{(0)}$ can be neglected, we obtain 

\begin{equation}
  \label{eq:t}
  t''=\sqrt{\frac{2m}{qfF_0\alpha}}.
\end{equation}

The important role of $t''$ manifests itself in the scaling laws for the Brunel-related and injection-current-related harmonics. After some straightforward but tedious calculations (see Appendix A for details), we obtain 
\begin{equation}
  \label{eq:Jm}
  \dot J_m^{\mathrm{br}}\propto t'', \dot J_m^{\mathrm{inj}} \propto \mathrm{const},
\end{equation}
where $\dot J_m^{\mathrm{br}}$ and $\dot J_m^{\mathrm{inj}}$  are the contributions to $m$th harmonics
from the Brunel and  injection current mechanisms correspondingly.

Taking into account \refeq{eq:Er} and the expressions for $t''$ above,
we conclude that the contribution to THz intensity from the Brunel
mechanism is proportional to $\alpha^{-1}$ if the initial velocity is negligible
and proportional to $\alpha^{-2}$ for the opposite case. On the other hand, the
contribution from the injection current does not depend on $\alpha$ in
the first approximation. These findings provide an important insight which allows  
to address and characterize the state of the electron immediately after ionization, as determined from $r^{(0)}$ and $v^{(0)}$. 
In the next section we test these scalings using direct numerical simulations of quantum equations.

\section{Numerical Simulations}

\begin{figure*}[tbph!]
  \centering
  \includegraphics[width=\textwidth]{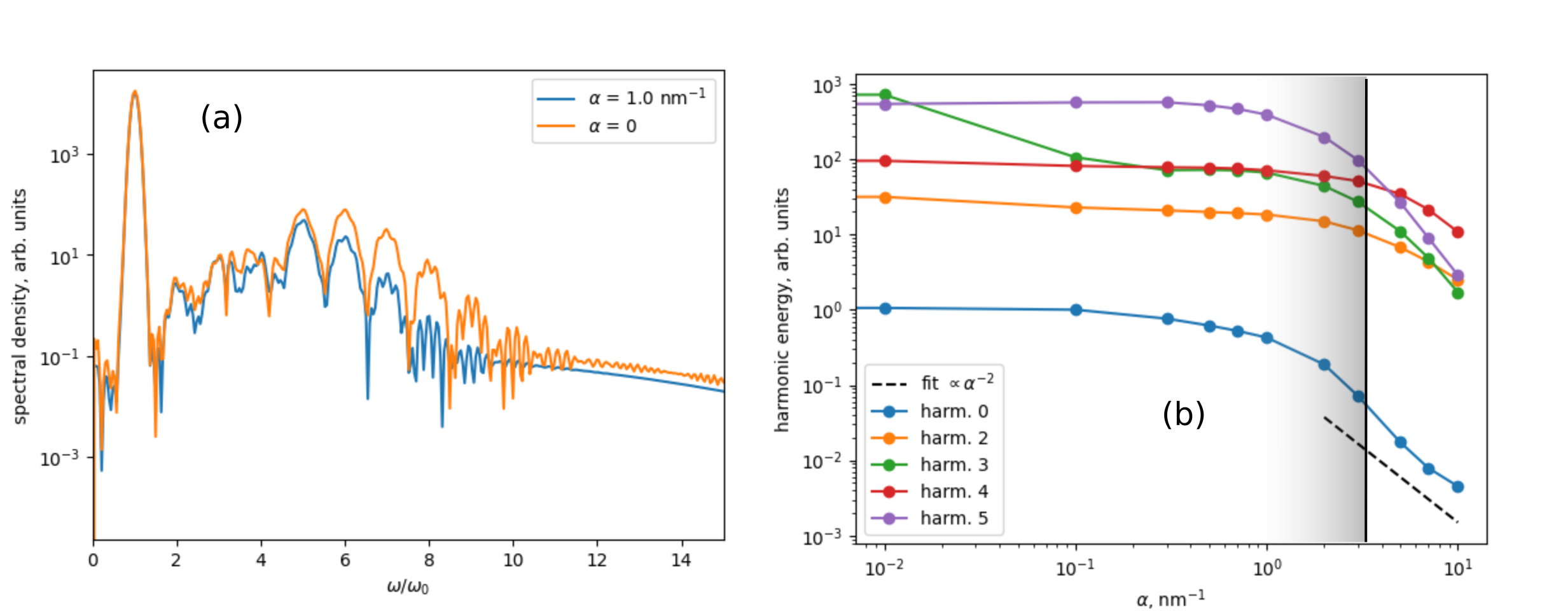}
  \caption{(a) Exemplary spectra of $E_r(\omega)$ according to TDSE
    simulations for few values of the field gradient $\alpha$.
    (b) The scaling of different harmonics with $\alpha$. Black dashed
  line shows an exemplary fit line. Vertical black line shows
  $\alpha_{\mathrm{bound}} = 1/x_\mathrm{max}$ according to
  \refeq{eq:xmax} (that is, for $v^{(0)}=0$). Shading indicates
  schematically the
  region of $\alpha\approx 1/x_\mathrm{max}$ for $v^{(0)}$ deviating
  from zero.}
  \label{fig:tdse}
\end{figure*}

To test our simple analytical theory above, we considered TDSE in the Coulomb gauge ($\partial_x A=0$):
\begin{equation}
  \label{eq:tdse}
  i\hbar\frac{\partial \psi(x,t)}{\partial t } =
\frac{1}{2m}\left[ (p + eA(x,t))^2 + V(x)\right]
  \psi(x,t)
\end{equation}
where $\psi(x,t)$ is the electronic wavefunction,
$p=-i\hbar\frac{\partial}{\partial x}$, $A(x,t)$ is the vector
potential defined in the same way as \refeq{eq:field-grad} with
$f=100$ and varying $\alpha$.
$V(x)$ is a rectangular asymmetric potential 
\begin{equation}
  \label{eq:pot}
V(x)=\begin{cases}
      -V_0,&\text{if $-2a<x<0$;}\\
      0,&\text{if $x>0$;}\\
      \infty,&\text{if $x<-2a$,}
    \end{cases}
\end{equation}
where $a=0.106$ nm, $V_0=16.94$ eV are selected in such a way that i)
the potential has exactly one bound state and ii) the ionization
potential of this bound state equals to the work function of gold (5.1
eV). We note that this potential assumes that the wavefunction inside
the metal is localized. The experimentally
observed electron spectra
\cite{kruger2011-nanotip-tunnnel,kruger12-metal-nanotip-rev}
are well-described by this
rather simple approach. Therefore, it is widely used to
model the ionization \cite{kruger12-metal-nanotip-rev,vampa17,dombi20,shi21,kiss22}. 
In the potential defined by \refeq{eq:pot}, ionization can occur only
in the positive direction of $x$. The electrons leaving the emitter
are accelerated by the field. Radiation was calculated by 
\refeq{eq:Er}, with $g=1$ for simplicity, and
\begin{equation}
  \label{eq:jquant}
  J = \langle \psi | p | \psi \rangle.
\end{equation}
The simulation was made by a split-step
method, with separate evaluation of the terms $\sim p^2$, $pA + Ap$,
$A^2$ and $V$; the action of $p$ was calculated using the fast Fourier
transform.

The results of simulations for the driving pulse with the Gaussian
shape, duration of 7 fs, and amplitude $fF_0=0.053$ a.u. (corresponding to 100 TW/cm$^2$) is shown in Fig.
\ref{fig:tdse}. In Fig. \ref{fig:tdse}(a) the spectra of $E_r(\omega)$ are shown for few exemplary
values of the field gradient $\alpha$, whereas in Fig. \ref{fig:tdse}(b) the
dependence of the intensity of radiated harmonics on $\alpha$ is
presented. $1/x_\mathrm{max}$ for $v^{(0)}=0$ is indicated by the vertical line.
As the simple semiclassical consideration above has predicted,
for small $\alpha\ll 1/x_\mathrm{max}$, there is almost no dependence of the
radiated harmonics on $\alpha$, with a notable exception of the third harmonic. However, as $\alpha$ approaches
$1/x_\mathrm{max}$, the harmonics intensity start to decrease, with
the scaling approaching $\approx 1/\alpha^{2}$ for 0th harmonic and
similar scaling for
higher harmonics.
Comparison of these results with the simple theory established in Section 3 indicates that, first,
radiating pattern corresponds to the dominance of the Brunel mechanisms
in this case. Second, the predicted scaling corresponds to a significant initial velocity $v^{(0)}$ of the
electron wavepacket at the tunneling exit. These findings highlight the differences of the photocurrent emission in the NS case and in the case of bulk crystal, where injection current dominates the emission \cite{jurgens20,jurgens24}. Note, however, that these results
are indirect indicators which should be used with some care and should be augmented by experimental studies (see Conclusions
for more details). 

We note that our semi-analytical analysis performed in the
approximation of ``infinitely sharp'' ionization steps should provide
increasingly bad estimation for the higher harmonics, specially as
the period of the corresponding light approach the ionization step
duration. In this respect, it is not appropriate to consider very high
harmonics (above around 7th) using the semi-analytical analysis. In particular, we see that for
5th harmonic the scaling
starts to be visibly higher than $\approx 1/\alpha^{-2}$, which can,
in particular, be an indication of an increasing role of return-to-the-ion (three-step)
harmonics. 

\section{Discussion and Conclusions }

In conclusion, here we predict that the Brunel and injection current
mechanism of harmonic emission in strong optical fields manifest
themselves not only in gases but also in metallic NSs. The essential
difference to the corresponding process in solids and gases originates
from significant local field gradients which is typical for NS. As we
showed, the field inhomogeneities start to play a notable role when
the maximal excursion of the electric field $x_\mathrm{max}$ is of the
order or larger than the inverse field gradient $1/\alpha$. In this
case, the electron trajectories are significantly modified by the
field gradients, leading to the decrease in the harmonic emission.
In this sense, studying the photocurrent-induced harmonics
  can bring information about the local field gradients. Whereas the
  photo-emitted electrons are routinely detected \cite{kruger2011-nanotip-tunnnel,kruger12-metal-nanotip-rev,dombi13,dombi20} and can provide the
  information about the local field gradients \cite{dombi13}, they
  might be not accessible in certain cases, for instance if the
  nanostructure is located inside a solid or liquid. 

Besides, photocurrent-induced harmonics are emitted 
  by electron in process of its leaving the NS, thereby imprinting the electron
dynamics \textit{in situ}. This is in contrast to imaging with electrons,
which, to deliver information about the immediate vicinity of
their birth in continuum, must be somehow ``back-propagated'' \cite{ni16}
from the position of their detection. To illustrate this,
we have developed a semiclassical model which allows us to derive the
scaling of the harmonic emission as a function of the field
inhomogeneity $\alpha$. We establish that this scaling allows an
additional insight into the generation mechanism and in particular
into the dynamics of the electrons near the tunnel barrier exit,
namely the electron velocity at the barrier exit. In particular, in
our TDSE simulations we obtained, that for large field gradients  the
generation efficiency decreases approximately as $\alpha^{-2}$, which
indicates nonzero velocity of the electronic wavepacket at the moment
of the tunneling to the continuum. Although in our numerical
  simulations we scanned $\alpha$, the  quantity
which matters is $x_\mathrm{max}\alpha$. It can be changed not only by
modifying $\alpha$ (which would be difficult experimentally), but also
by changing intensity and wavelength (cf. Eq. 10).

This is consistent with the recent studies for tunneling in atoms,
suggesting nonzero electron velocity at the tunnel exit
\cite{camus17}. Here we note that that the electron wavepacket is
localized neither in momentum nor in space, so one should speak about
the initial electron velocity at the barrier exit with certain care.
This equally applies to the tunneling time, that is the time, which
the electron spends under the barrier, making it heavily dependent on
the measurement procedure \cite{landsman15,zimmermann16,sokolovski18}.
The same must be true also for the initial electron velocity. In our
case, as mentioned above, this is the velocity which the radiated
harmonics bear a fingerprint of. It must not even be identical for
every particular harmonic, in the same sense as the tunneling time
seen by particular recollision-based harmonic can in principle depend
on the number of that harmonic \cite{zhao13}. In view of these
unclarified questions and conceptual uncertainties it is of particular
importance to design new characterization approaches for the tunneling
process and pertinent electron dynamics, such as the one proposed in
our paper. We note that the efficiency of the
  photocurrent-induced harmonics could be rather low; besides, they can be
  shadowed by other nonlinear mechanisms, such as strong nonlinearities
  in metals \cite{shi19,boyd14,babushkin23}. This might emerge as a challenge in the future experimental realizations. Yet, we believe
  that at least for the zeroth harmonic and with optimized
  nanostructure geometry these
  problems can be overcome.

We conclude that the photoinduced current-based
harmonics can be generated by electron emission in NSs, not only in gases and
  solids. They can be furthermore used as a tool,  allowing to "look inside" the
initial stages of the photoemisison dynamics, and the NSs with their strong field
gradients provide additional possibilities for that.

\appendix

\section*{Appendix A. Semiclassical calculation of scaling at $x_{\mathrm{max}}\gg
  \alpha$}
We recast the piecewise-dependent electric field as given by \refeq{eq:e-simpl} in the following form:
\begin{equation}
  \label{eq:e-simpl2}
  E(t,x) = \Theta(t,t',t''(x))f E_0(t) + \Theta(t,t''(x),\infty)E_0(t),  
\end{equation}
with $t'$ being the electron birth time and $t''(x)$ being the time
moment at which the
electron achieves the position $x=x''\equiv 1/\alpha$ in the homogeneous field
$E(t,x) = E_0(t)$. Here $\Theta(t,t_1,t_2)$ is a step function:
$\Theta(t,t_1,t_2)$ is one of $t_1<t<t_2$ and zero
otherwise.

Equipped with the  definitions above and taking into account the definitions
of the piece-wise functions $\Theta(t,t_1,t_2)$ we can replace \refeq{eq:1el} with
(assuming here both the field and velocity to be scalar): 
\begin{gather}
  \nonumber
  m\dot v  = -qE_0(t)\left(f\Theta(t,t',t'') -
  \Theta(t,t'',\infty)\right)\\
  =m(\dot v)_1 - m(\dot v)_2,
  \label{eq:1el1}
\end{gather}
where $(\dot v)_i$ are defined as $m(\dot v)_i=-qf_i\Theta(t-t_i')E_0(t)$
with $f_1=1$, $f_2=f-1$, $t_1=t'$, $t_1=t'$, $t_2=t''$.

The meaning of \refeq{eq:1el1} is the following: 
electron acceleration can be represented a sum of two contribution of the type
\refeq{eq:1el}, first is from the electron born at the time $t'$ and
propagating in the homogeneous field $fE_0(t))$, and the second, with the
opposite sign, from the electron born at time $t''$ and propagating in the
homogeneous field $(f-1)E_0(t)$. For $t\ge t''$, we have both terms
present. If $f\gg 1$ and therefore
\begin{equation}
  \label{eq:f}
  f-1\approx f,
\end{equation}
these terms have approximately equal amplitudes but opposite signs, and
therefore cancel each other. That is, we can represent the switching
off the field by spurious creation of an ``electron with opposite
charge sign'' (one can imagine this as the creation of a positron),
which contribution for the time $t>t''$ cancel the contribution of the
already existing electron. Note that cancellation takes place for if we
consider acceleration $\dot v$, but not for $v$ itself. Yet, for the
radiation, only $\dot v$ is important.

To apply this argument to the net growth rate of the current
 $\dot J$ in \refeq{eq:dj} (here $\dot J $ is the scalar version of $d\vect J/dt$ for our
one-dimensional model), we note that, because
ionization rate $\dot \rho$ is the very nonlinear function of the
field strength, the electrons are born mostly in the vicinity of the
maxima of electric field $t_n$. Because of asymmetry of our
structure, only the maxima with definite sign of $E_0$, for instance,
$E_0<0$, contribute to the ionization rate. The ionization 
therefore takes place
in sharp steps of the typical width of order of hundred attoseconds. For this reason, as
soon as we consider lowest harmonics with the oscillation period still
in the femtosecond range, we can represent $\dot \rho$ as a sum of the delta-functions
and $\rho$ as sum of steps:
\begin{gather}
      \label{eq:steps-rho}
  \dot \rho = \sum_n \delta\rho_n \left(\delta(t-t_n) -
    \delta(t-t_n-t'')\right), \\ \rho =
  \sum_n\delta\rho_n\left(\Theta(t-t_n) - \Theta(t-t_n-t'')\right),
    \label{eq:steps-w}
\end{gather}
where $\delta(t)$ is the Dirac $\delta$-function, $\delta \rho_n$ is
amplitude of the $n$th step, $t''$ is given by \refeq{eq:t} and corresponds to the ``spurious
positron'' creation, which should model the inhomogeneity of the field
by effectively switching it off, introduced as described above. In
\refeq{eq:steps-rho} we silently assumed \refeq{eq:f}, that is, that
the field away from the NS is negligible, which makes step
sizes by positive and negative terms in \refeq{eq:steps-rho} equal,
that is, negative terms completely cancel positive terms for
$t>t_n+t''$.

Substituting \refeq{eq:steps-rho} to \refeq{eq:dj}, we
obtain:
\begin{gather}
\label{eq:steps-j-sum}
  \dot J = \sum_n\left(\dot J_n(t) - \dot J''_n(t)\right),  \\
  \nonumber
  \dot
  J_n(t)=-\frac{2U\delta\rho_n}
      {|E(t_n)|}\dot \delta(t-t_n)   \\
  + \delta\rho_n\frac{2q^2}{m} E(t_n)\Theta(t-t_n),
  \label{eq:steps-j}
\end{gather}
and the expression for $\dot J_n''(t)$ is obtained from the expression
for $\dot J_n(t)$ by replacing $t_n$ with $t_n+t''$. 

Now we take into account that for any (sufficiently smooth) function $g(t)$ we have
\begin{gather}
\nonumber
g(t_n)\Theta(t-t_n)-g(t+t'')\Theta(t-t_n-t'') = \\
\nonumber
  g(t_n)\Theta(t,t_n,t_n+t'') - \left(g_n't'' +
    g_n''t''^2/2+\ldots\right)\\
  \times\Theta(t-t_n-t''),
  \label{eq:theta-g-expansion}  
\end{gather}
where we used the Tailor expansion of $g(t)$ and denoted $g_n'$,
$g_n''$ the derivatives of $g$ at $t=t_n$.  Analogously,
\begin{gather}
\nonumber
g(t_n)\dot \delta(t-t_n)-g(t_n+t'')\dot \delta(t-t_n-t'') =\\
\nonumber
  g(t_n)\dot\delta(t,t_n,t_n+t'') - \left(g_n't'' +
    g_n''t''^2/2+\ldots\right)\\
  \times\dot\delta(t-t_n-t''),
  \label{eq:delta-g-expansion}  
\end{gather}
where $\dot\delta(t,t_n,t_n+t'')$ is defined as
$\dot\delta(t,t_n,t_n+t'')=\dot\delta(t-t_n)-\dot \delta(t-t_n-t'')$.
Here we assume $g(t)=2\delta \rho_nq^2E(t)/m$ for the case with
$\Theta$ and $g(t)=-2U\delta \rho_n/|E(t)|$ for the case with $\dot
\delta$ (cf. \refeq{eq:steps-j}).  We note that since the electrons
are released near the extrema of the electric field, $g'_n=0$, 
only the terms with $g''_n$ (and higher) remain in the Tailor expansion
in \refeq{eq:theta-g-expansion} and \refeq{eq:delta-g-expansion}. 

Equipped with expressions above, we can  analyze the dependence of the
harmonics on $\alpha$ in the region $x_{\mathrm{max}}\ll
\alpha$. From the contributions from different terms to the harmonics
given by $\dot J_m = \mathcal{F}[\dot J(t)](m\omega_0)\propto\int \dot J
e^{-im\omega_0t}dt$ we obtain that the contribution of the term
$\Theta(t,t_n,t_n+t'')$ is proportional to $t''$, the
contribution of the term with $\Theta(t-t_n-t'')$ is proportional to
$g_n''t''^2/(m\omega_0)$, whereas the contributions from the terms with
$\dot \delta(t,t_n,t_n+t'')$ and $\dot \delta(t-t_n-t'')$ are
proportional to  $g_n''t''^2/(m\omega_0)$ and  $1/(m\omega_0)$. 
Keeping the leading terms in $t''$, we obtain \refeq{eq:Jm}.  

Note that the scaling with $t''$, calculated above for every single ionization event $t_n$, is readily transferred to the whole pulse. Indeed, as suggested by \refeq{eq:steps-j-sum}, we must sum over all ionization events $t_n$. By this summation, the constructive  interference takes place \cite{babushkin11} at frequencies $\omega=m\omega_0$, recovering the standard  appearance of harmonics in the emission. On the other hand, since the scaling mentioned above is valid for every particular ionization event, it is readily translated to every harmonic as a whole by the summation. 

\section*{Acknowledgements}
I.B., A.D., M.K and U.M acknowledge  support by the Cluster of Excellence PhoenixD (Photonics, Optics, and Engineering-Innovation Across Disciplines), DFG EXC 2181/1. A. H. acknowledges support from DFG, project 1593/16-1.








\begin{thebibliography}{9}

\bibitem{vampa17} Vampa, G., \textit{et al}.,
   "Attosecond nanophotonics,"
  Nat. Photonics vol.~11, p.~210, 2017.
  
\bibitem{dombi20} Dombi, P \textit{et al}.,
   "Strong-field nano-optics,"
  Rev. Mod. Phys. vol.~92, p.~025003, 2020. 

  \bibitem{kruger2011-nanotip-tunnnel}
M. Kr\"uger, M. Schenk, and P. Hommelhoff, "Attosecond control of electrons emitted from a nanoscale metal tip," Nature,
vol. 475, p.~78-81, 2011.

\bibitem{kruger12-metal-nanotip-rev}  M. Kr\"uger, M. Schenk, M.
  F\"orster, P. Hommelhoff,  "Attosecond physics in photoemission from a metal nanotip," J. Phys. B: Atom.,
    Molec. Opt. Phys., vol. 45, p.~074006, 2012.

\bibitem{dombi13} P. Dombi, \textit{et al.},  "Ultrafast
 strong-field photoemission from plasmonic nanoparticles." Nano
 lett. vol.~13, p.~674, 2013. 

    
\bibitem{ludwig2019sub}
M.~Ludwig, G.~Aguirregabiria, F.~Ritzkowsky, T.~Rybka, D.~C. Marinica,
  J.~Aizpurua, A.~G. Borisov, A.~Leitenstorfer, D.~Brida, "Sub-femtosecond electron transport in a nanoscale gap," Nature Phys. vol.~16, p.~341, 2020.

\bibitem{shi21} L. Shi, \textit{et al}.
  "Femtosecond field-driven on-chip unidirectional electronic
  currents in nonadiabatic tunnelling regime,"
  Laser \& Photonics Rev. vol.~15, p.~2000475 (2021).

  
\bibitem{schoetz2019perspective}
J.~Schoetz, Z.~Wang, E.~Pisanty, M.~Lewenstein, M.~F. Kling, M.~Ciappina, "Perspective on Petahertz Electronics and Attosecond Nanoscopy,"
ACS Photonics, vol.~6, p.~3057,  2019.



  \bibitem{schiffrin13}
A.~Schiffrin, T.~Paasch-Colberg, N.~Karpowicz, V.~Apalkov, D.~Gerster,
    S.~M{\"u}hlbrandt, M.~Korbman, J.~Reichert, M.~Schultze, S.~Holzner
    \emph{et~al.},  "Optical-field-induced current in dielectrics,"
    Nature, vol.~493, p.~70, 2013.
  
\bibitem{karnetzky18}
C.~Karnetzky, P.~Zimmermann, C.~Trummer, C.~Duque~Sierra, M.~W\"orle,
    R.~Kienberger, and A.~Holleitner,  "Towards femtosecond on-chip
    electronics based on plasmonic hot electron nano-emitters,"
    Nat. Commun., vol.~9, p.~2471, 2018.

    
  \bibitem{sem} Taryl L. Kirk,
"A Review of Scanning Electron Microscopy in Near Field Emission Mode,"
Adv. Imag. Electron Phys., vol.~204, p.~39, 2017.


  \bibitem{plasm_em} F. Schertz, M. Schmelzeisen, M. Kreiter, H.-J. Elmers, and G. Schönhense, 
"Field Emission of Electrons Generated by the Near Field of Strongly Coupled Plasmons," Phys. Rev. Lett., vol.~108, p.~237602, 2012.

\bibitem{ciappina14} M. F. Ciappina, \textit{et al.}
  "High-order-harmonic generation driven by metal nanotip
  photoemission: Theory and simulations." Phys. Rev. A
  vol.~89,p.~013409, 2014.

\bibitem{kim08} Kim, S. \textit{et al}.
    "High-harmonic generation by resonant plasmon field enhancement,"
   Nature, vol.~453, p.~757, 2008.

\bibitem{brunel90cp}
F.~Brunel,  "Harmonic generation due to plasma effects in a gas undergoing multiphoton ionization in the high-intensity limit,"
    J. Opt. Soc. Am. B, vol.~ 7, pp.~521--526, 1990.

\bibitem{kim08b}
K.~Y. Kim, A.~J. Taylor, J.~H. Glownia, and G.~Rodriguez,  "Coherent control of terahertz supercontinuum generation in ultrafast laser-gas interactions," Nat. Photonics, vol. 2, pp.~605--609, 2008.

\bibitem{zhang17} Zhang, X. C. et al., "Extreme terahertz science," Nature Photon., vol.~11, p.~16, 2017.

\bibitem{koulouklidis20} Koulouklidis, A. D. \emph{et al.}, "Observation of extremely efficient terahertz generation from mid-infrared two-color laser filaments," Nat. Commun., vol.~ 11, p.~292, 2020.

\bibitem{babushkin11}
I.~Babushkin, S.~Skupin, A.~Husakou, C.~K\"{o}hler, E.~Cabrera-Granado,
    L.~Berg\'{e}, and J.~Herrmann,  "Tailoring terahertz radiation by controlling tunnel photoionization events in gases,"
    New J. Phys., vol.~13, p.~123029, 2011.

\bibitem{babushkin17}
I.~Babushkin, C.~Br{\'e}e, C.~M. Dietrich, A.~Demircan, U.~Morgner, and A.~Husakou,  "Terahertz and higher-order Brunel harmonics: from tunnel to multiphoton ionization regime in tailored fields," J. Mod. Opt., vol. 64, pp.~1078--1087, 2017.

\bibitem{lanin17} Lanin, A., \textit{et al}., "Mapping the electron band structure by intraband
high-harmonic generation in solids," Optica, vol.~4, p.~516, 2017.

\bibitem{jurgens20} Ju\"rgens, P. \textit{et al}., "Origin of
  strong-field-induced low-order harmonic generation in amorphous
  quartz," Nature Phys., vol.~16, p.~1035, 2020.

\bibitem{babushkin22} I. Babushkin, \textit{et al}.,
  "All-optical attoclock for imaging tunnelling wavepackets," Nature
  Phys., vol.~18, p.~417, 2022.

\bibitem{shi20} L.~Shi, \textit{et al.}, "Progressive self-boosting
  anapole-enhanced deep-ultraviolet third harmonic during few-cycle
  laser radiation." ACS Photon. vol.~7 p.~1655, 2020.

\bibitem{jurgens24} P. Juergens {it et al.}, "Linking High-Harmonic Generation and Strong-Field Ionization in Bulk Crystals," ACS Photonics, vol.~11, p.~247, 2024. 

\bibitem{kiss22}  G. Zs Kiss, P. F\:oldi, and P. Dombi. "Ultrafast
  plasmonic photoemission in the single-cycle and few-cycle regimes."
  Sci. Rep. vol.~12, p.~3932, 2022.
  
\bibitem{hhg_np}
A. Husakou, S.-J. Im, and J. Herrmann, "Theory of plasmon-enhanced high-order harmonic generation in the vicinity of metal nanostructures in noble gases,"
Phys. Rev. A, vol.~83, p.~043839, 2011.

\bibitem{ni16} H. Ni \textit{et al.}, ``Tunneling ionization time
  resolved by backpropagation,'' Phys. Rev. Lett., vol.~117,
  p.~023002, 2016.

\bibitem{camus17}
N. Camus, E. Yakaboylu, L. Fechner, M. Klaiber,
M. Laux, Y. Mi, K. Z. Hatsagortsyan, T. Pfeifer,
Ch. H. Keitel, and R. Moshammer, ``Experimental Evidence for Quantum Tunneling Time,''
Phys. Rev. Lett., vol.~119, p.~023201, 2017.

\bibitem{landsman15}
Landsman, A. S. and Keller, U. ``Attosecond science and the tunnelling
time problem,'' Phys. Rep. vol.~547, pp.~1–24, 2015.
  
\bibitem{zimmermann16}
Zimmermann, T., Mishra, S., Doran, B. R., Gordon, D. F. and Landsman, A. S., "Tunneling time and weak measurement in strong field ionization," Phys. Rev. Lett., vol.~117, p.~023002, 2016.

\bibitem{sokolovski18}
D. Sokolovski and E. Akhmatskaya, 
"No time at the end of the tunnel,"
Commun. Phys. vol.~1, p.~47, 2018.
    
\bibitem{zhao13}
J. Zhao and M. Lein, "Determination of Ionization and Tunneling Times in High-Order Harmonic Generation," Phys. Rev. Lett., vol.~ 111, p.~043901, 2013.

\bibitem{shi19} L. Shi, \textit{et al.} "Generating ultrabroadband
  deep-UV radiation and sub-10 nm gap by hybrid-morphology gold
  antennas." Nano lett. vol.~19, p.~4779, 2019.  

\bibitem{boyd14} R. W. Boyd, Z. Shi, and I. De Leon, ``The third-order nonlinear
optical susceptibility of gold,'' Opt. Commun. vol.~326, p.~74, 2014.

\bibitem{babushkin23} I. Babushkin, \textit{et al.} "Metallic
  nanostructures as electronic billiards for nonlinear terahertz
  photonics." Phys. Rev. Research vol.~5, p.~043151, 2023.

\end{thebibliography}
\end{document}